\documentstyle[12pt]{article}
\begin{document}

\setlength{\baselineskip}{24.2pt}

\newcommand\ie {{\it i.e. }}
\newcommand\eg {{\it e.g. }}
\newcommand\etc{{\it etc. }}
\newcommand\cf {{\it cf.  }}
\newcommand\etal {{\it et al. }}
\newcommand{\be}{\begin{eqnarray}}
\newcommand{\ee}{\end{eqnarray}}
\newcommand\Jpsi{{J/\psi}}
\newcommand\M{M_{Q \overline Q}}
\newcommand\mpmm{{\mu^+ \mu^-}}
\newcommand{\jp}{$ J/ \psi $}
\newcommand{\pp}{$ \psi^{ \prime} $}
\newcommand{\ppp}{$ \psi^{ \prime \prime } $}
\newcommand{\dd}[2]{$ #1 \overline #2 $}
\newcommand\noi {\noindent}

\begin{center}
{\bf Rapidity Distributions of Dileptons from a
Hadronizing Quark-Gluon Plasma}\\[5ex]

R. Vogt$^1$, B. V. Jacak$^2$, P. L. McGaughey$^2$, and P. V.
Ruuskanen$^3$\\[3ex]

$^1$Gesellschaft f\"{u}r Schwerionenforschung (GSI), D-6100 Darmstadt 11,
Germany\\[2ex]

$^2$Los Alamos National Laboratory, Los Alamos, NM 87545, USA\\[2ex]

$^3$Research Institute for Theoretical Physics, University of Helsinki,\\
P.O. Box 9, FIN-00014 Helsinki, Finland\\
and\\
Department of Physics, University of Jyv\"{a}skyl\"{a},\\
P.O. Box 35 FIN-40351 Jyv\"{a}skyl\"{a}, Finland\\[2ex]

\end{center}

\vspace{1cm}
\begin{center}
{\bf Abstract}
\end{center}
\begin{quote}
\begin{small}
It has been predicted that dilepton production may be used as a quark-gluon
plasma probe.  We calculate the rapidity distributions of thermal dileptons
produced by an evolving quark-gluon plasma assuming a longitudinal scaling
expansion with initial conditions locally determined from the hadronic
rapidity density.
These distributions are compared with Drell-Yan
production and semileptonic charm decays at invariant mass $M = 2$, 4, and 6
GeV.
\end{small}
\end{quote}
\newpage

One of the challenges in ultrarelativistic heavy ion physics is to
differentiate the thermal signatures of a quark-gluon plasma
from particles produced in the initial nucleon-nucleon interactions.
In order for dilepton production to be a useful plasma probe one must
then distinguish between
thermal production and the significant `background' dilepton production from
initially produced Drell-Yan and charm pairs.
We focus on the rapidity
distributions of lepton pairs in the mass region between $2 \leq M \leq 6$ GeV
where thermal
production could be competitive with initial dilepton production.  We also show
that thermal charm production is significant.

The rapidity distribution of Drell-Yan pairs of invariant mass $M$ in
central (impact parameter $b=0$) $AA$
interactions, averaged over protons and neutrons, is defined as
\be \frac{dN}{dMdy} & = & \frac{A^{4/3}}{\pi r_0^2} K \frac{8\pi \alpha^2}{9
M^3} \left[ \frac{5}{18}
( x_a v(x_a) x_b s(x_b) + x_a s(x_a) x_b v(x_b) + 4 x_a s(x_a) x_b s(x_b))
\right. \\ \nonumber &   & \hbox{} + \left. \frac{4}{18} x_a s(x_a) x_b s(x_b)
\right] \, \, , \ee
where $K=2$ accounts for higher-order corrections, $v(x)= u_v(x) +
d_v(x)$ is the valence quark distribution, and
$s(x) = 2(\overline u(x) + \overline
d(x) + \overline s(x))/6$ is the average sea quark distribution in a nucleon
assuming a flavor symmetric sea.
We have used the new lowest-order set of structure functions, Duke-Owens 1.1
\cite{Owens}, in our calculations.
The fractions of the initial projectile and target momenta carried
by the interacting partons are given by $x_a = M e^y/\sqrt{s}$ and $x_b =
Me^{-y}/\sqrt{s}$, respectively.

The rapidity distributions from Drell-Yan production in a central Au+Au
collision
are shown in Fig.\ 1(a) for
RHIC energies and 1(b) for the LHC.  At RHIC, the rate rises
slightly for $M = 2$ GeV and stays fairly constant over three units of rapidity
even for $M = 6$ GeV.  It drops off steeply at the kinematic limit.  At the
LHC, the rate rises somewhat over four units of rapidity for all masses.  This
effect is most prominent at the LHC for $M = 2$ GeV where $x_b \leq 0.00032$
and $x_a \geq 0.00032$. The sea-sea term in eq.\ (1) is the dominant
contribution and increases slightly since $x_b$ remains near the small $x$
limit of the structure functions while $x_a$ increases with $y$.  The
first sea-valence contribution in eq.\ (1) increases with $y$ but the second
decreases because a small decrease in $x_b$ leads to a fast decrease of $x_b
v(x_b)$.  Thus the total Drell-Yan rate increases with $y$ by
$\approx$ 30 percent.

The charmed hadron production cross section in a nucleon-nucleon
collision is \cite{VBH2,Ban} \be E_3 E_4 \frac{d \sigma}{d^3p_3 d^3p_4} & = &
K \int \frac{\hat{s}}{2 \pi} \frac{dx_a}{x_a} \frac{dx_b}{x_b} dz_3 dz_4
H(x_a,x_b) \frac{E_3 E_4}{E_1 E_2} \\ \nonumber &   &
\frac{D(z_3)D(z_4)}{z_3^3 z_4^3}
\delta^4 (p_a + p_b - p_1 - p_2) \, \, , \ee where subscripts $a$ and $b$
represent the incoming partons, 1 and 2, the charmed quarks, and 3 and
4, the charmed hadrons.  The function $D(z)$ describes the fragmentation of
the charmed quarks into hadrons.  In ref.\ \cite{VBH2} it was shown that
charm hadroproduction is best described by the assumption that
the charmed quark loses no momentum during hadronization, {\it i.e.}\ $D(z) =
\delta(1-z)$.  Given this, {\it e.g.}\ $p_{\parallel_1} = p_{\parallel_3}$,
but $y_1 \neq y_3$.  (A fragmentation function assuming some momentum
loss results in a more central rapidity distribution.)
The convolution function $H(x_a,x_b)$ is \be
H(x_a,x_b) = \sum_{n_f} (x_a q_{n_f}(x_a) x_b\overline q_{n_f}(x_b) +
x_a \overline q_{n_f}(x_a) x_bq_{n_f}(x_b))
\frac{d \hat{\sigma}}{d \hat{t}}|_{q \overline q} + x_ag(x_a) x_bg(x_b)
\frac{d \hat{\sigma}}{d \hat{t}}|_{gg} \, \, , \ee where $q(x)$ and $g(x)$ are
the parton distributions \cite{Owens}.  For the explicit
subprocess cross sections, see {\it e.g.}\ ref.\ \cite{Ellis}.

We assume no
intrinsic $p_T$ for the light quarks and use four-momentum conservation to
integrate over $x_a$ and $x_b$.  Then the number of charmed hadrons
produced in a central $AA$ collision is
\be \frac{d N}{dM_{D \overline D}
dy_{D \overline D}} & = & \frac{A^{4 \alpha/3}}{\pi r_0^2} 2 K \int
dp_T^2 dz_3 dz_4 dy_3 dy_4 M_{D \overline D} H(x_a, x_b)
\frac{E_3 E_4}{E_1 E_2}
\frac{D(z_3) D(z_4)}{z_3 z_4} \nonumber \\
&   & \delta(y_{D \overline D} - (y_3 + y_4)/2) \delta(M^2_{D \overline D} -
(p_3 + p_4)^2) \, \, , \ee
where $M_{D \overline D}$ and $y_{D \overline D}$ are the mass and
rapidity of the $D \overline D$ pair.
The uncertainty in the magnitude of the charm production cross
section from a lowest-order calculation results in $K \sim 2-3$.
The shape of the rapidity
distribution is relatively unaffected by next-to-leading order corrections
\cite{Nason}.
Since $pA$ measurements \cite{Pat,Appel,E769}
indicate $\alpha \sim
0.9-1$, we use $\alpha = 1$ to give an upper
bound on initial charm production.  Using eq.\ (4), we can
determine the rapidity distribution of lepton pairs from the decay
$D \overline  D \rightarrow l^+ l^- + {\rm anything}$
and from the average branching ratio $B(D/D^{\star}
\rightarrow l^+ + {\rm anything}) \sim$ 12\%.
($B(D^0 \rightarrow l^+ + {\rm anything})
=$ 7.7\% and $B(D^+ \rightarrow l^+ + {\rm anything}) =$ 19.3\% while
$D^{0 {\star}} \rightarrow D^0 X$ and $D^{+ {\star}} \rightarrow
D^+/D^0 X$ where $X$ is either a photon or pion.)
Given a $D \overline D$ pair with mass and rapidity distributed according
to eq.\ (4), the decay to dileptons is performed. The pair is first assigned
a transverse momentum and polar angle using a parametrization of the charm
pair correlation data of ref.\ \cite{Koda}. Then the two $D$'s are
semileptonically decayed according to the lepton energy spectrum reported in
ref.\ \cite{Balt}.
At high energies and for large nuclei more than one $D \overline D$ pair may be
produced so that the detected
charm decay leptons may be uncorrelated.

The charmed hadron rapidity distributions are shown in Figs.\ 2(a) and (b) for
Au+Au collisions at RHIC and LHC (solid curves) along with the resulting
lepton pair
distributions at $M=2$ (dashed), 4 (dot dashed), and 6 (dotted) GeV
(scaled up by 100 to fit on the same plot).  Charm decays are copious at
these energies ($\sigma_{D \overline D} \sim 190 \, \mu$b at RHIC and 1.95 mb
at
LHC using $m_c = 1.5$ GeV and $K = 2$, compatible with the next-to-leading
order
results \cite{Nason}).  Measurements suggest that charm decays are important
in the dilepton spectrum at $M = 2$ GeV at SPS energies \cite{London,Carlos}.
Fermilab experiment
E789 at 800 GeV observed that the dilepton continuum at $M < M_{J/\psi}$
is much greater than expected from Drell-Yan alone, indicating that the
dominant background may be due to charm decays \cite{Mike}.
A judicious choice of kinematic cuts can reduce the
charm acceptance relative to that of Drell-Yan.
Drell-Yan pairs are produced with equal but
opposite transverse momentum with $p_T<M$ while the transverse momenta of
the leptons
from charm decays are only weakly correlated.  This is discussed in more detail
below.

We have not included shadowing of the nuclear structure functions in our
calculation.  Shadowing would lead to a depletion of both the charm and
Drell-Yan cross sections in the very low $x_b$ region.
The effect is strongest at small $y$ since both the beam and target
distributions are shadowed, leading to a faster increase
of the Drell-Yan rate with $y$.  Aside from the uncertainties in the rates due
to shadowing,
it is important to note that the nucleon structure functions have
not yet been measured in the LHC $x$ region and are also not well known for
RHIC.  HERA will study the region $10^{-5} \leq x$ and $Q^2 > 10$ GeV$^2$.
Since extrapolations of the measured structure functions to the low $x$ region
differ widely, the HERA results could, by favoring some models over
others, lead to significant changes in the calculations of
$\sigma_{\rm DY}$ and $\sigma_{D \overline D}$.  For example, using
KMRS B0 \cite{KMRS}, a higher-order
set of structure functions with a larger sea quark
contribution at low $x$,
increases the Drell-Yan cross section by a factor of two.  Early results
from HERA favor sea quark and gluon distributions that increase as
$x \rightarrow 0$, suggesting that, particularly at
LHC energies, QCD calculations are quite sensitive to
the choice of structure functions \cite{ep}.

Thus far we have been considering
the expected background to the thermal dileptons,
now we discuss the signal itself.  At RHIC and LHC it is reasonable to
expect the longitudinal expansion to follow the scaling law, $v_z=z/t$, so that
the initial density is related to the final multiplicity distribution by $n_i
\tau_i = (dN/d\eta)/(\pi R_A^2)$.  The temperature is therefore a function of
$\eta$, $T_i \equiv T_i(\eta)$.
The multiplicity distribution is parametrized
as \be \frac{dN}{d\eta} = \left(\frac{dN}{d\eta}\right)_0
\exp(-\eta^2/2\sigma^2) \, \, , \ee where $(dN/d\eta)_0$ is the total
multiplicity at $\eta =0$ and $\sigma$ is the width.  At RHIC
energies, $|\eta| \leq 6$ and we choose $(dN/d\eta)_0 = 2000$ while for the
LHC $|\eta| \leq 8$ and we take $(dN/d\eta)_0 = 5000$.  Assuming that the
Gaussian shape extends to the kinematical limit, the energy sum rule is
saturated for $\sigma\sim 3$ at RHIC and $\sim 5$ at LHC.  This
form as well as the scaling properties of the
longitudinal velocity must, of course, break down at rapidities close to the
phase space limit \cite{Kat2} where the densities become too small for
thermal production to be observed.


The dilepton emission rate is \cite{KR,OP}
\be \frac{dN}{d^4x d^2p_T dy dM^2} = \frac{\alpha^2}{8\pi^4} F \exp{ [-(m_T
\cosh(y - \eta)/T)]} \, \, , \ee where $M$, $p_T$, $m_T$, and $y$ are the
mass, transverse momentum, transverse mass, and rapidity of the pair and $\eta$
is the rapidity of the medium with temperature $T$.  In a plasma, $F_Q = \sum
e_q^2$, while in an equilibrium hadron gas, assuming $\pi \pi \rightarrow
\rho \rightarrow
l^+ l^-$ is the dominant channel, $F_H = \frac{1}{12} m_\rho^4 /((m_\rho^2 -
M^2)^2 + m_\rho^2 \Gamma_\rho^2)$.  The mixed phase has a fractional
contribution from each. After integration over $p_T$, $\tau$, and transverse
area,
the rapidity distribution for a given dilepton mass is
\be &~& \frac{dN}{dM dy} = \frac{\alpha^2}{2 \pi^3} \pi R_A^2 \left\{ \theta
(s_i - s_Q) \frac{3 F_Q }{M^3} \int_{-Y}^{Y}
d\eta (\tau_i T_i^3)^2 \frac{x^4 + 5x^3 + 15x^2 + 30x + 30}{\cosh^6(\eta - y)}
e^{-x}|_{x_c}^{x_i} \right. \nonumber \\
&~& \mbox{} + \theta(s_i - s_H) M^3 \frac{\tau_m^2}{2} f_0 (r-1) \left[f_0
F_Q + ((r-2)f_0 + 2)F_H \right] \int_{-Y}^{Y} d\eta \left( \frac{1}{x_c} +
\frac{1}{x_c^2} \right)e^{-x_c} \\
&~& \mbox{} + \left. \theta(s_i - s_{\rm dec}) \frac{3 F_H}{M^3} \int_{-Y}^{Y}
d\eta (\tau_H T_H^3)^2 \frac{x^4 + 5x^3 + 15x^2 + 30x +
30}{\cosh^6(\eta - y)} e^{-x}|_{x_{\rm dec}}^{x_H} \right\} \, \, , \nonumber
\ee where $x = M \cosh(\eta - y)/T$, $s_i$ is the  initial entropy density and
$s_Q$ and $s_H$ are the entropy densities of the  plasma and the hadron gas at
the transition temperature $T_c$.

We neglect the transverse expansion and describe the properties of the matter
in the longitudinal direction only. Transverse expansion mainly affects the
hadron gas phase and the later part of the mixed phase \cite{PVR,Kat1}.
If the transverse
expansion is included, the contribution from the hadron gas phase is negligible
for $M\geq 2$ GeV except at large
rapidities where the initial density is too low to produce a plasma.

As defaults, we assume a three-flavor plasma and a hadron gas of massless
pions with $T_c = 200$ MeV and $T_{\rm dec} = 140$ MeV.  Then \[ s_i = \left\{
\begin{array}{ll} 4 \gamma_k \frac{\pi^2}{90}T_i^3 & \mbox{$s_i > s_Q$ or $s_i
< s_H$} \\ f_0 s_Q + (1-f_0)s_H & s_H < s_i < s_Q \end{array} \right. \]
where $\gamma_k$ is the number of degrees of freedom with
$\gamma_H = 3$ and $\gamma_Q = 16 + 21n_f/2$ and $n_i = s_i/3.6$.
The beginning of the mixed phase occurs at \[
\tau_m = \left\{ \begin{array}{ll} \tau_i (T_i/T_c)^3 & T_i > T_c \\ 1 \, {\rm
fm} & T_i = T_c \, \, . \end{array} \right. \]  The initial plasma content of
the mixed phase is \[ f_0 = \left\{ \begin{array}{ll} 1 & T_i > T_c \\ (s_i
- s_H)/(s_Q - s_H) & T_i = T_c \, \, . \\ 0 & T_i < T_c \end{array} \right. \]
The beginning of the hadron phase is then \[
\tau_H = \left\{ \begin{array}{ll} \tau_m r & T_i \geq T_c \, \, ,
\\ 1 \, {\rm fm} & T_i < T_c  \end{array} \right. \]
where $r = s_Q/s_H = \gamma_Q/\gamma_H$.
There is no thermal contribution if $T_i < T_{\rm dec}$.

We have used three different hypotheses to fix $T_i$ and $\tau_i$.  First we
assume a fixed initial time, $\tau_i \equiv 1$ fm.  We then define
$dN/d\eta|_Q = \pi R_A^2 n_Q \tau_i$, $dN/d\eta|_H = \pi R_A^2 n_H \tau_i$, and
$dN/d\eta|_{\rm dec} = \pi R_A^2 n_{\rm dec} \tau_i$.
If $dN/d\eta > dN/d\eta|_Q$ the
system begins in the plasma phase and \be T_i = \left\{ \frac{dN}{d\eta}
\frac{T_c^3}{\pi R_A^2 n_Q \tau_i} \right\}^{1/3} \, \, . \ee  If $dN/d\eta|_H
< dN/d\eta < dN/d\eta|_Q$, the system begins in the mixed phase.  If
$dN/d\eta|_{\rm dec} < dN/d\eta < dN/d\eta|_H$, it begins in the hadron gas
with \be T_i = T_H = \left\{
\frac{dN}{d\eta} \frac{T_c^3}{\pi R_A^2 n_H \tau_i} \right\}^{1/3} \, \, . \ee
There is no contribution if $dN/d\eta < dN/d\eta|_{\rm dec}$.  The
maximum initial temperature, $T_{i,{\rm max}}$, obtained with a fixed initial
time is the lowest of all three hypotheses.
In this case we find $T_{i,{\rm max}} \sim 255$ MeV at RHIC and 350 MeV at the
LHC.

As an alternative way of determining the initial time $\tau_i$ and
temperature $T_i$, we use the bound $\tau_i\langle E_i\rangle \simeq
3\tau_iT_i=C\geq 1$ from the uncertainty relation. We consider two cases, $C=3$
and $C=1$, where the latter gives an absolute lower bound on $\tau_i$
and an upper bound on $T_i$.  Resent results from the
parton cascade model \cite{Geiger} and estimates of gluon thermalization
\cite{Shuryak} give some support for a short formation time. In this case, the
initial temperature of the plasma is
\be T_i = \left\{ \frac{dN}{d\eta}
\frac{3T_c^3}{\pi R_A^2 n_Q C} \right\}^{1/2} \, \, . \ee  If $T_i \leq T_c$,
then the formation time is again fixed to be $\tau_i = 1$ fm and the system
begins in the mixed phase or hadron gas, according to the multiplicity, as
before. When $C=3$, we find $T_{i,{\rm max}} \sim 300$ MeV at RHIC and 465 MeV
at the LHC.  Using $C = 1$, we have $T_{i,{\rm max}} \sim 515$ and 810 MeV
respectively, similar to the results of ref.\ \cite{Joe}.

In Fig.\ 3 we show thermal dilepton production at RHIC for $M = 2$ GeV (a) and
(b), $M=4$ GeV (c) and (d), and $M=6$ GeV (e) and (f).  The full evolution
(plasma, mixed phase, and hadron gas) is shown in (a), (c), and (e) while the
plasma contribution alone is given in (b), (d), and (f). The thermal dilepton
distributions arising from the fixed initial time with $T_i$ from eq.\ (8)
(solid) and the uncertainty bound from eq.\ (10) with $C = 3$ (dashed) and $C =
1$ (dot-dashed) are given. Dilepton production by the plasma is most important
at central rapidities where the initial temperature is greatest.  Including the
mixed phase and the hadron gas broadens the rapidity distributions. Without
transverse expansion, the lifetime of the hadron gas phase at central
rapidities, where the initial density is high, becomes very long (of order 80
fm when $T_i \sim T_{i,{\rm max}}$) and its contribution to the thermal pair
rate $dN/dMdy$ is thus unphysically large. At higher rapidities, the calculated
hadron contribution is more physical since the initial density is too low to
form a plasma and the effects of transverse expansion remain small. In Fig.\
3(a), (c), and (e) the thermal dilepton distributions merge at higher
rapidities since $\tau_i \equiv 1$ fm when $T_i \leq T_c$. Later time
contributions are less important when $C = 1$ because of the high $T_{i,{\rm
max}}$. The distributions from plasma alone, Fig.\ 3(b), (d), and (f), are
narrower and have a reduced yield.  Since it is difficult to produce high mass
pairs when $T_i \leq T_c$, at $M = 6$ GeV the yield at $y=0$ is unchanged in
Fig.\ 3(e) and (f). The results for the LHC are given in Fig.\ 4.
Due to the  increased $T_{i,{\rm max}}$, the contribution from the plasma
dominates the thermal emission for $M > 2$ GeV at this energy.
To show the most
optimistic results of thermal dilepton production, we will compare the $C=1$
curves with the other contributions to the dilepton spectrum.  One must keep in
mind, however, that the other choices of initial conditions result in a much
smaller thermal yield for $M > 2$ GeV.

We have also varied our inputs to see how the rapidity distributions are
affected.  Increasing the hadron degrees of freedom reduces the mixed phase and
hadron gas contributions since their lifetimes are decreased.
Changing to a two-flavor plasma increases the initial temperature considerably.
For $\tau_i \equiv 1$, $T_{i,{\rm max}}$ grows to 285 MeV at RHIC and 386 MeV
at the LHC.  When $C = 3$, we have 340 MeV and 540 MeV while when $C = 1$, we
find 590 and 930 MeV respectively.  Setting $T_c = 140$ MeV and $T_{\rm dec} =
100$ MeV increases $\tau_m$ but results in a decrease of the mixed phase and
hadron contributions due to the larger $x_c$ and $x_{\rm dec}$ (lower $T_c$ and
$T_{\rm dec}$) in eq.\ (7).

If $C=1$ represents the more physical situation, one might
expect a significant amount of thermal charm
production since $T_{i,{\rm max}} \sim (1/3 - 1/2) m_c$. The importance
of thermal charm production has previously been suggested \cite{Shuryak,Shor}.
For completeness, we also estimate the rapidity distribution of dileptons
from thermal charm production.  The thermal $c \overline c$ production rate
is \be \frac{dN}{d^4 x} = \int \frac{d^3 k_a}{(2\pi)^3} f(k_a)\frac{d^3
k_b}{(2\pi)^3} f(k_b) \sigma_{c \overline c}(M) v_{\rm rel} \, \, , \ee
where $v_{\rm rel}$ is the relative velocity of the thermal partons,
$v_{\rm rel} = \sqrt{(k_a \cdot k_b)^2 - m_q^4}/(E_a E_b)$.  We use a
Boltzmann distribution for the incoming partons, as before.  Following ref.\
\cite{Shor}, we write the charm production cross section as \be
\sigma_{c \overline c}(M) = \gamma_q \sigma_{q \overline q \rightarrow c
\overline c}(M) + \gamma_g \sigma_{gg \rightarrow c \overline c}(M) \, \, , \ee
where the quark and gluon degeneracy factors are $\gamma_q = 3 \times (2 \times
3)^2$ for three quark flavors and $\gamma_g = (2 \times 8)^2/2$ (the factor of
1/2 is to prevent double counting) \cite{Shor}.  The cross sections are given
in ref.\ \cite{Comb}.  We consider only production from the plasma phase for
this case since the later stages of evolution do
not have sufficiently high temperatures to provide significant thermal charm
production.  We have not included charm fragmentation
since this has only a minor effect on the rapidity
distribution as compared to that of the source.  Therefore,
we may simply write \be \frac{dN}{dM dy} =
\frac{3 \sigma_{c \overline c}(M)}{4 M (2\pi)^4} \pi R_A^2 \theta
(s_i - s_Q) \int_{-Y}^{Y}
d\eta (\tau_i T_i^3)^2 \frac{x^4 + 5x^3 + 15x^2 + 30x + 30}{\cosh^6(\eta - y)}
e^{-x}|_{x_c}^{x_i} \, \, . \ee  The resulting distributions are similar in
shape to the thermal dileptons but broader than the initial charm production.
Thermal charm production is comparable to that in the initial
nucleon-nucleon interactions.  In Figs.\ 2(c) and (d) we compare the
thermal and initial charm production.  The mass distribution is much steeper
for the thermal production, resulting in a smaller decay contribution at higher
masses.  For comparison, at RHIC, the $M=2$ GeV curve is multiplied by a factor
of 1000 while the $M=4$ and 6 GeV results are multiplied by $10^4$ and $10^5$
respectively.  At the LHC, the
multipliers are 100 for $M=2$ GeV, 1000 for $M = 4$, and $10^4$ for $M = 6$
GeV.

Our combined results for lepton pairs with $M = 2$, 4, and 6 GeV are given
in Figs.\ 5 and 6 at RHIC and
LHC respectively.  The Drell-Yan and initial charm contributions
are given by
the dashed and dot-dashed curves.  The thermal dileptons arising from direct
production over the full time evolution and charm decays produced by plasma
only with $C=1$ are given by the solid and dotted
curves.
The thermal dileptons have the broadest
distribution of all the contributions at RHIC while the decays of
initially produced charm have the
narrowest.  The slopes are also somewhat different near the central region.
At $M=2$ GeV, the thermal contributions are above the Drell-Yan.  At the LHC,
the thermal rate remains compatible with the Drell-Yan yield
and all the slopes are similar.  However, at $y>3$ the thermal and charm decay
contributions are decreasing while the Drell-Yan rate is still increasing.

Although the initial charm contribution is very large, it may be possible to
separate it from the thermal dilepton and Drell-Yan contributions.  At RHIC $e
\mu$ coincidence measurements could prove useful.  Charm was first measured by
this method at the ISR \cite{Chil} and such coincidence measurements are
planned at RHIC \cite{CDR}. Opposite sign $e \mu$ pairs from decay of two
charmed particles are correlated if only one $D \overline D$ pair is produced
in an event, which is the case for $pp$ and light ion collisions at RHIC.  If
$\alpha = 1$ in eq.\ (4), up to five primary $D \overline D$ pairs could be
produced in Au+Au collisions, yielding uncorrelated $e \mu$ pairs. The
production rate of Drell-Yan and thermal dilepton pairs is much less than one
pair per event so uncorrelated $e \mu$ production from these processes is very
unlikely. However, thermal charm may be important if the thermalization time is
short, and cause significant additional production of uncorrelated $e \mu$
pairs. Pairs of like-sign electrons may offer a better measure of charm
production if more than two $D \overline D$ pairs are produced per event. This
is probable for all ion species in the LHC environment, where $e \mu$
coincidence measurements will not be possible in any case \cite{Alice}.

Figs. 5 and 6 suggest a very large correction to the measured dilepton
distributions if subtraction of the charm contribution is attempted. However,
these figures do not include the experimental acceptance, smaller
for leptons from charm decay than for Drell-Yan or thermal dileptons.
Table 1 shows the acceptances in the muon channel
of the RHIC PHENIX detector at $M=2$, 4, and 6 GeV and $y=2$.
The Drell-Yan and thermal acceptances are from four to twenty times
larger than for the charm decay leptons \cite{CDR}. Consequently,
the thermal and Drell-Yan contributions do not represent only a few
percent of the total dilepton yield and subtraction of the charm
contribution may thus be feasible. This is illustrated in Table 2
which shows the ratios of the cross sections without, then
with the effects of acceptance.
It is important to note that in the case of dileptons from
uncorrelated charm decays, the acceptance drops by another factor of nearly
two due to the larger rapidity difference between the leptons.
It should be possible to optimize a rapidity cut to further
discriminate against dileptons from uncorrelated charm. Whether such a
cut can be used to reject a majority of the combinatorial lepton pairs
from uncorrelated charm can only be addressed for specific detector
geometries.

If it is possible to reliably remove the charm decay contributions to the
dileptons, the significant differences between the rapidity
distributions of thermal and Drell-Yan dileptons could be measured.
The Drell-Yan contribution grows slightly with $y$ while the thermals
decrease with $y$.  The correlation of thermal production with the hadronic
multiplicity and the effect of shadowing on Drell-Yan
should lead to opposite trends in
the mass number dependence of these two contributions. Stronger shadowing for
higher mass nuclei favors the larger $y$ region for Drell-Yan pairs.
Increasing the mass number should narrow the rapidity distribution
due to the stronger stopping and consequently also narrow
the thermal dilepton distribution.
The shape of the distributions in the central several units of rapidity should
thus provide an indication of the dilepton source. A significantly broad
rapidity and mass coverage at RHIC and LHC would allow such
a study.

In summary, we have calculated the rapidity distributions of dilepton
production from various sources in heavy ion collisions at RHIC and LHC.
The calculated distributions cover four or more units of rapidity, suggesting
the contributions to dilepton measurements over a wide range of rapidities.
A large experimental
rapidity coverage is desirable to allow a search for variations in the shape of
the dilepton rapidity distributions in $pp$, $pA$ and $AA$ collisions.
Charmed particle decays into leptons are very important,
but the kinematics of the charm decays differ from that of thermal or
Drell-Yan dilepton production. Consequently the acceptance of realistic
detectors makes the charm background which must be subtracted
to study the thermal dileptons more tractable.
Furthermore the significant production of thermal
charm may offer a novel way to study the collisions, since the rate is
comparable to direct thermal dileptons. Experiments could search for
broadening of the dilepton distribution in $AA$ collisions after fixing
the level of charm production in $pp$ interactions.

We would like to thank S. Gavin and G. Young for useful discussions.
RV thanks Los Alamos National Laboratory and the University of
Jyv\"{a}skyl\"{a} for hospitality.

\newpage
\begin{center}
{\bf Tables}
\end{center}

\vspace{0.5in}

\begin{table}
\begin{center}
\begin{tabular}{|c|c|c|c|} \hline
Mass & 2 GeV & 4 GeV & 6 GeV \\ \hline \hline
Drell-Yan 	  & 3.9\% & 5.4\% & 4.6\% \\ \hline
Thermal dileptons & 4.9   & 6.8   & 5.8 \\ \hline
charm decays 	  & 0.88  & 0.89  & 0.25 \\ \hline
\end{tabular}
\end{center}
\caption[]{Acceptance in PHENIX muon arm for dileptons from various
sources at $y=2$.}
\end{table}

\vspace{0.5in}

\begin{table}
\begin{center}
\begin{tabular}{|c|c|c|c|}
\multicolumn{4}{c}{Ratio of dilepton source to Drell-Yan}  \\ \hline
Mass & 2 GeV & 4 GeV & 6 GeV \\ \hline \hline
Thermal (C=1) & 6.1  & 0.25 & 0.013 \\ \hline
Charm 	      & 66.  & 33.  & 3.8   \\ \hline
Thermal Charm & 14.  & 0.73 & 0. \\ \hline
\multicolumn{4}{c}{Acceptance correction included} \\ \hline
Mass & 2 GeV & 4 GeV & 6 GeV \\ \hline \hline
Thermal (C=1) & 7.7  & 0.31 & 0.016 \\ \hline
Charm 	      & 15.  & 5.5  & 0.21  \\ \hline
Thermal Charm & 3.1  & 0.12 & 0. \\ \hline
\end{tabular}
\end{center}
\caption[]{The upper section gives the ratio of dileptons produced by other
sources to Drell-Yan production at RHIC.  The lower section shows
the effect of the PHENIX muon acceptance on these ratios at $y=2$.}
\end{table}

\vspace{2.5in}

\newpage

\newpage
\begin{center}
{\bf Figure Captions}
\end{center}
\vspace{0.5in}
Figure 1.  Rapidity distributions of Drell-Yan pairs at $M = 2$ (solid),
4 (dashed), and 6 (dot dashed) GeV in a central Au+Au collision at RHIC (a) and
the LHC (b).\\

\vspace{0.2in}

\noi Figure 2.  Initially produced charmed pair distributions (solid curves)
from a central Au+Au collision at RHIC (a) and at the LHC (b).  Also shown are
the lepton pair distributions resulting from semileptonic decays of charm at
$M_{D \overline D} = 2$ (dashed), 4 (dot dashed), and 6 (dotted) GeV.  The
lepton pairs from the decays have been scaled up by a factor of 100.  Thermal
charm production is shown in (c) at RHIC and (d) at the LHC.  The decay leptons
from thermal charm at RHIC are scaled up by a factor of 1000 for $M=2$ GeV
(dashed), $10^4$ at $M=4$ GeV (dot dashed), and $10^5$ at $M=6$ GeV (dotted).
At the LHC, the decays with
$M=2$ GeV are scaled up by 100 while those with $M=4$ (dot dashed) and $M=6$
GeV (dotted) are scaled up by 1000 and 10000.\\

\vspace{0.2in}

\noi Figure 3.  The rapidity distribution of thermal dileptons produced at RHIC
with $\sigma = 3$ in eq.\ (5) are shown for (a) and (b) $M = 2$ GeV, (c) and
(d) $M = 4$ GeV, and (e) and (f) $M = 6$ GeV. The thermal curves represent the
three choices of $\tau_i$ and $T_i$, $\tau_i \equiv 1$ fm (solid), $C = 3$
(dashed), and $C = 1$ (dot dashed).  The full evolution (plasma, mixed phase,
and hadron gas) is given in (a), (c), and (e) while the plasma contribution
alone is shown in (b), (d), and (f).\\

\vspace{0.2in}

\noi Figure 4.  The same as in Fig.\ 3 for the LHC with $\sigma = 5$.\\

\vspace{0.2in}

\noi Figure 5.  We compare the contributions to the dilepton spectrum at RHIC
for (a) $M =2$ GeV, (b) $M=4$ GeV, and (c) $M=6$ GeV.  Drell-Yan (dashed),
direct thermal dileptons with $C=1$ (solid), thermal (dotted) and initial (dot
dashed) charm production are included.\\

\vspace{0.2in}

\noi Figure 6.  The same as in Fig.\ 5 for the LHC.\\

\vspace{0.2in}

\end{document}